\documentstyle[twocolumn]{mn} 
\input{epsf.tex}
\def\beqa{\begin{eqnarray}}
\def\eeqa{\end{eqnarray}}
\def\beq{\begin{equation}}
\def\eeq{\end{equation}}
\def\beqn{\begin{displaymath}}
\def\eeqn{\end{displaymath}}
\def\beqan{\begin{eqnarray*}}
\def\eeqan{\end{eqnarray*}}

\def\hm{\ \rm {\it h}^{-1} Mpc}
\begin{document} 

\title{Imprints of Primordial Voids on the CMB}  
\author[C.Baccigalupi, L. Amendola and F. Occhionero]
{C. Baccigalupi$^{1,2}$, L. Amendola$^{1}$ and F. Occhionero$^{1}$\\
$^{1}$ Osservatorio Astronomico di Roma, 
Viale del Parco Mellini 84, 00136 Roma, Italy\\  
$^{2}$ Department of Physics, University of Ferrara,
Via del Paradiso 12, 44100 Ferrara, Italy}

\maketitle 
\baselineskip14pt
  
\begin{abstract} 
We generalize in several ways the results existing in the literature:
a) we make use of an exact general relativistic solution for a spherical,
nearly empty cavity in the matter dominated era to evaluate the null 
geodesics and the Sachs-Wolfe effect; b) we evaluate the magnitude of 
the adiabatic fluctuations of the photon-baryon plasma; c) we study 
the influence of the shell profile; and d) we take into account the 
finite thickness of the last scattering surface (LSS) and the influence 
of its position with respect to the void center. We find empirically
an analytic approximation to the Sachs-Wolfe
effect for all crossing geometries and we derive an upper limit of 
$\approx$ 25 $h^{-1}$ Mpc for the comoving radii of voids sitting on
the LSS in order to achieve compatibility with COBE's
data. As a nearly empty void has an overcomoving
expansion of a factor of $ \approx$ 4 between decoupling and the present,
the maximum allowed size at present is $\approx$ 100 $h^{-1}$ Mpc. 
On the other hand, the smallness of the comoving  size relative to 
the sound horizon reduces strongly the adiabatic effect   by Silk 
damping and makes it negligible.  Most of the signature of primordial voids 
comes therefore from metric effects and consists of subdegree spots blue or 
red depending on whether the center lies beyond or within the LSS. 
In conclusion we refine and confirm earlier constraints on 
a power law void spectrum originated in an inflationary phase transition
and capable of generating the observed large scale structure.

\end{abstract} 
 
\section{introduction} 
 
Inflation provides two well known mechanisms 
for relating primordial physics to perturbations on 
large scales ($\ge 10 \hm$) in the observed 
universe. 
The first one arises from the small quantum fluctuations 
of the inflaton field: the fluctuations are stretched out to 
scales greater than the horizon ($H^{-1}$) by  the
spatial expansion, and their dynamics freezes until the 
reenter in post-inflationary epochs. The second mechanism 
occurs in the case of a first order phase transition in the early
universe; the nucleated bubbles imprint
the density field. In fact, in this work we assume that they are the 
primordial cause of the large scale voids 
observed at the present. A class of inflationary models provides bubbles 
production; the  
underlying minimal physics is given by quantum field theory  
through quantum transition of  the inflaton field from a false vacuum 
state to a true vacuum one. 
More realistic models of first order inflation (FOI), 
after the original version by Guth (1981), were
found by considering non-minimal 
gravity theories and two-field phenomenology (see e.g. La \&  
Steinhardt 1989, Kolb 1991, Adams \& Freese 1991, Occhionero \& Amendola 
1994), and made FOI the richest extension of the slow-rolling 
model worked out by Linde (1983).

The hypothesis of a
bubbly distribution of matter in the universe, 
theoretically interesting in itself, is strongly suggested by
recent observations. Large-scale voids 
($\ge 10\,\hm$ ) have been discovered by recent redshift 
surveys (see e.g. Kauffmann \& Fairall 1991, Vogeley, 
Geller \& Huchra 1991, Lindner et al. 1995, and references therein), 
but the definition of
a statistics for their dimensions is still premature, 
due to the presence of voids comparable with the sample's size. 
Deep ($1000 \hm$ ) pencil-beam surveys toward the galactic poles 
revealed a fascinating regularity in the galaxy distribution with
a characteristic scale of 128 $\hm$  (Broadhurst et al. 1990); 
El-Ad et al. (1996) constructed an algorithm to find voids in redshift
surveys; da Costa et al. (1996) found that the voids are really empty 
of matter.

The estimate of the anisotropies induced by a void in the
cosmic microwave background (CMB) yields the strongest 
constraint to the underlying
inflationary physics
by comparison with COBE observations
(see e.g. Bennet et al. 1994 and references therein).
Detections of degree scale CMB anisotropies (see e.g. Scott et al.
1996) provide in our case weaker 
constraints because they involve a limited part of the sky. 
To be quantitative, upgrading 
the experiment from 10 to 1 degree  would imply a gain of a factor 100
in resolution; but in our case we would be looking for one single object, 
the largest, 
and we have to take into account the portion of sky in which we search
(precisely rescaling by the corrisponding factor $dN_{B}$ in Eq.(37) below).
Consequently, such observations would be significative in our
case if the covered solid angle were at least 400 square degrees, and
to our knowledge high resolution observations of this width do not exist.

The first attempt to relate the evidence of voids in the large scale 
matter distribution with FOI was made by La in 1991. Overcoming the
difficulties of previous works (Turner et al. 1992, 
Liddle \& Wands 1991, 1992), it has been found that it is possible
to construct models that fit
the galaxy correlation function and pass the CMB bounds
with a bubbly matter distribution  
(Amendola \& Occhionero 1993, Amendola \& Borgani 1994). 
Occhionero \& Amendola (1994) found a toy model of FOI where the expected 
 distribution in radius of the primordial voids, 
defined as the number of voids with radius greater than 
$R$ today,  is approximated by a power law,
\begin{equation}
N_{B}(R)=\left({R_{M}\over R}\right)^{p}\ \ .
\label{distribution}
\end{equation}
This distribution is compatible with the
measured galaxy power spectrum and with the
CMB anisotropies for 
\begin{equation}
R_{M}\simeq 30\,\hm\,  \ \ ,\ \ 6<p<10\ \ ,
\label{values}
\end{equation}
normalizing $N_B$ to a cube with $500 \hm$ by side.
Aim of this paper is to review and to set the constraint on $R_{M}$ 
and $p$ more rigorously than in the original paper, by an accurate
evaluation of the physics involved.

A void is characterized by its central 
to asymptotical density ratio  
\begin{equation} 
\Delta=\frac{\rho_{center}}{\rho_{\infty}} 
\label{Delta} 
\end{equation} 
that decreases with time. 
The dynamics of voids has been investigated by Icke (1984)
and  Bertschinger (1983, 1985)  
in the strongly non-linear regime, $\Delta\ll 1$, and for 
$t \rightarrow \infty$. The important
result of their analysis is that 
the physical radius $r$ of a compensated void of total energy $E$ 
increases as $t^{4/5}$ and
that all the significative quantities of a void 
(pressure, density profile, velocity  
field, mass) are only functions of the dimensionless coordinate 
$\lambda=r(Et^{2}/\rho_{\infty})^{-1/5}$, thus remaining unchanged 
at late times (self-similar expansion). These analyses have been 
confirmed by $N$-body simulations (White \& Ostriker 1990,  
Dubinski et al. 1993). 
In the following we mean that $R$ in Eq. (\ref{distribution}) 
is the physical radius today; however, the corrisponding comoving
quantity was smaller in the past by a factor of  
$(t/t_{0})^{{4\over 5}-{2\over 3}}=(1+z)^{-0.2}$;
in particular, a void on the last scattering
surface (LSS), at $1+z\simeq 1000$, was physically smaller by a 
factor of $4 \times 1000$.

If we connect present voids with primordial bubbles,
the energy profile at nucleation time can in principle
be calculated exactly by solving the inhomogeneous 
Klein-Gordon equation for the inflaton field with the 
appropriate tunneling-like initial conditions. The general 
structure is a spherically symmetric underdensity, with a 
thin wall, and a size of the order of $H^{-1}$ (Coleman 1977).
The internal matter content depends on the detailed shape
of the transition potential.
However, its successive evolution depends crucially
on the properties of the  dominating
cosmic fluid. From Vadas (1993) we know that 
super-horizon voids, during the radiation dominated era (RDE),
undergo hydrodynamical inflow of the primordial plasma
into the empty central region. For an estimate of this effect
we call $\eta$ the mean velocity dispersion 
of the primordial plasma in units $c=1$, where the cases $\eta=0,1$ 
mean respectively predominance and absence of CDM; simply, 
we sketch the filling process as inflow of cosmic fluid in a spherical 
empty void of comoving size $R/4$ (the 4 factor accounting for the 
overcomoving expansion) from 
the end of inflation ($T_{e}\simeq 10^{15} \rm{GeV}$) across equivalence 
($T_{EQ}\simeq 10 \rm{eV}$) until decoupling ($T_{D}\simeq 2\cdot 10^{-1}
\rm{ eV}$);  
one finds that the quantity $\Delta$ defined in (\ref{Delta}) is given by
\begin{equation} 
\Delta (t_{D})=1-\exp\left[-\eta{2H_{o}^{-1}\over R/4}
\sqrt{T_{o}\over T_{D}}
\left(3-\sqrt{T_{D}\over T_{EQ}}\right)\right]\ . 
\label{filling}
\end{equation}
For $R\simeq 10\hm$ our 
estimate gives $\Delta\simeq 1$ for $\eta=O({1\over 100})$, a very typical
value for CDM (La 1991). Thus, 
voids smaller than $R\simeq 10 \hm$ are erased by matter inflow;
this is therefore a lower cut-off in the distribution (\ref{distribution}).
We can see that larger
voids ($ R\simeq 10 \div 50 \hm)$  have a typical filling fraction 
$.5\ge\Delta\ge .1$ in their central underdense region.
Since no significative dependence of our results upon these values 
exists, in the following we will assume $\Delta\simeq 0.3$.

We use exact metric models, built by
Occhionero et al. (1981,1983),  
for the description of spherical, compensated structures directly in the  
matter dominated era (MDE). We investigate 
the metric distortion induced on the CMB radiation by the voids
and check our approach by comparison with previous analyses of 
strongly non-linear highly approximated voids completely crossed
by photons (Thompson \& Vishniac 1987, hereafter TV).
CMB complete crossing of voids was treated recently by some authors 
(Panek 1992, Arnau et al. 1993). 
We find that the metric CMB distortion in the case of complete
crossing, Rees-Sciama effect (1968, hereafter RS), is negligible with respect 
to partial crossing, Sachs-Wolfe effect (1967, hereafter SW), 
for voids intersecting the last scattering surface (LSS), 
the width of which is modelled by making use of exact decoupling 
physics. Therefore, for treating accurately this case, we evaluate the 
CMB adiabatic distortion before decoupling due to the void. 
This is achieved by adapting the 
formalism developed by Hu and Sugiyama (1995, hereafter HS) to our case: 
it is shown that 
on scales in which SW effect is already comparable with the detected CMB 
anisotropies, the adiabatic effect is significatively 
reduced by Silk damping and then may be neglected. 
These results have been obtained by recovering, both for the SW and 
the adiabatic case, the angular dependence of the CMB temperature, 
which is therefore a function of the  void size, of its exact 
density profile and of the particular position of the void with respect to
the LSS; our model responds to all these items.

We anticipate that the largest void on the LSS that does not
violate the COBE observations has a comoving size of $\simeq 25\hm
\, $, i.e. physical size 100$\hm$ by today due to over comoving growth.
  
The content of this paper is as follows. In Sec.II we construct null 
geodesics . In Sec.III 
we analyse the CMB distortion induced by complete crossing of 
a void, comparing with previous results based on highly
approximated models. Finally, in Sec.IV we consider the
observational consequences of the  voids lying on LSS. 
In the last section we draw the conclusions.
  
\section{Geodesics}
 
The most natural theoretical enviroment to describe
non-linear spherical large-scale perturbations was found by 
Tolman in 1934. We make the minimal assumption of pressureless 
fluid, according to the CDM paradigm.

The fundamental length element is ($c=1$)
\begin{eqnarray}
ds^{2} &=& dt^{2}-{1\over\Gamma^{2}(M)}
\left({\partial r\over\partial M}\right)^{2}
dM^{2}-\nonumber\\
&-& r^{2}(M,t)(\sin^{2}\theta d\phi^{2}+d\theta^{2})\ \ ,
\label{ds2}
\end{eqnarray}
where $M$, the gravitating mass, plays the role of a radial
coordinate, and $\Gamma(M)$, a generalization of the Lorentz
$\gamma$, is a conserved quantity in the pressureless case; 
it is given by
\begin{equation}
\Gamma^{2}(M)=1+\left({\partial r\over\partial t}\right)^{2}
-{2GM\over r}\ \ .
\label{gamma}
\end{equation}
Also
\begin{equation}
\rho(M,t)={1\over 4\pi r^{2}\partial r/\partial M}\ \ .
\label{ro}
\end{equation}
The $\Gamma(M)$ function characterizes the perturbation structure. 
The choice $\Gamma =1$ reproduces the unperturbed Friedmannian case.
For hyperbolic perturbations, we write
\begin{equation}
\Gamma^{2}(M)=1+\Gamma_{+}^{2}(M)\,,
\label{gamma+}
\end{equation}
and the solution to Eq. (\ref{gamma}) takes the form
\begin{eqnarray}
r(M,t)&=&{GM\over\Gamma_{+}^{2}(M)}
\left({\rm cosh}\eta-1\right)\label{solution}\\
t&=&{GM\over\Gamma_{+}^{3}(M)}
\left({\rm sinh}\eta-\eta\right)\ \ .
\label{solutiont}
\end{eqnarray}
The model that we use here for $\Gamma_{+}^{2}(M)$, Occhionero et al. 
(1981,1983), is suitable for description
of a wide class of compensated matter structures,
and has the following 
analytical form:
\begin{equation}
\Gamma_{+}^{2}(x)=B\left({L_{s}\over L}\right)x^{2/3}
\gamma_{n}(x)\ \ ,\ \ 
x={M\over M_{*}}\ \ ,
\label{modello}
\end{equation}
\begin{equation}
L_{s}=2GM_{*}\ \ ,\ \ L=\left({3M_{*}\over 4\pi\rho_{\infty,o}}\right )^{1/3}
\ \ ,\ \ \label{modellol} 
\end{equation} 
\begin{eqnarray} 
\gamma_{n}(x)&=&{\rm exp}\left[-\left(\Gamma_{E}(1/n)/n\right)^{n}
x^{5n/3}\right]\label{modellos}\\
(\Gamma_{E}&\equiv& {\rm \ Euler\ Function}).\nonumber
\end{eqnarray}
The model contains a tunable scale 
$M_{*}$ that will be 
typically $10^{15}{\rm M}_{0}$ for a large scale void.
The $BL_{s}/L$ factor is a measure of the perturbation strength.
The exponential in Eq.(\ref{modellos})
sets to zero the perturbation for $M\gg M_{*}$; 
$\gamma_{n}(x)$ becomes a sharper and sharper stepwise function
by increasing $n$, while simultaneously 
the compensating shell becomes  narrower and 
denser; the coefficient in the exponent is dictated by the request of 
measuring the energy associated with the perturbation. 
With this model for $\Gamma_{+}(M)$, the dynamics 
is obtained solving Eqs.(\ref{solution}),(\ref{solutiont}).

We have now a model of
spacetime; the next step is to integrate the null geodesics.
We place the observer on the direction $\phi=0$,
$\theta=\pi /2$ from the void's center.
A point on a null geodesic is marked by an affine parameter 
$l$, and the wave vector is
\begin{equation}
\left(K^{t},K^{M},K^{\theta}=0,K^{\phi}\right)={d\over dl}
\left(t,M,\theta={\pi\over 2},\phi\right)\ \ .
\label{wave}
\end{equation}
The geodesic equations are 
\begin{equation}
{dK^{t}\over dl}=-{1\over\Gamma^{2}}{\partial r\over\partial M}
{\partial^{2}r\over\partial t\partial M}\left(K^{M}\right)^{2}-
r{\partial r\over\partial t}\left(K^{\phi}\right)^{2}\ \ ,
\label{eq1}
\end{equation}
\begin{eqnarray}
{dK^{M}\over dl}&=&\left({\partial r\over\partial M}\right)^{-1}
r\Gamma^{2}\left(K^{\phi}\right)^{2}-{1\over\Gamma}
{\partial\Gamma\over\partial M}\left(K^{M}\right)^{2}-\nonumber\\
&-&2\left({\partial r\over\partial M}\right)^{-1}
{\partial^{2}r\over\partial t\partial M}K^{t}K^{M}-\nonumber\\
&-&\left({\partial r\over\partial M}\right)^{-1}
{\partial^{2}r\over\partial M^{2}}\left(K^{M}\right)^{2}\ \ ,
\label{eq2}
\end{eqnarray}
\begin{equation}
{dK^{\phi}\over dl}={-2K^{\phi}\over r}
\left({\partial r\over\partial M}
K^{M}+{\partial r\over\partial t}K^{t}\right)\ \ .
\label{eq3}
\end{equation}
Of these, Eq.(\ref{eq3}) integrates directly into
\begin{equation}
K^{\phi}={k\over r^{2}}\ \ ,
\label{k}
\end{equation}
where the constant $k$ is fixed by the initial conditions.
We now integrate numerically Eqs.(\ref{eq1})(\ref{eq2}) and (\ref{k})
using the normalzation
\begin{equation}
K^{\alpha}K_{\alpha}=0\ \ ,
\label{eq4}
\end{equation}
as a check for the numerical accuracy. We set the ``initial"
conditions at the observation point $P_{0}$ and we integrate backwards.
After choosing $(K^{t})_{0}=-1$, 
Eq. (\ref{eq4}) at $P_{0}$ reads
\begin{eqnarray}
(K^{M})_{0}&=&\left[-\Gamma\left({\partial r\over\partial M}
\right)^{-1}
\sqrt{(K^{t})^{2}-{k^{2}\over r^{2}}}\right]_{0}=\nonumber\\
&=&-\left({3M\over r}\right)_{0}{\rm cos}\alpha\ \ ,
\label{Kmo}
\end{eqnarray}
where in the last equality, obtained by imposing 
that the observation point lies in the Friedmann region, 
$\alpha$ is the angle that the
geodesic under investigation forms at $P_{0}$ with the direction of 
the void's center:
\begin{equation}
k=r_{0}{\rm sin}\alpha\ \ .
\label{kI}
\end{equation}
With the relations $K^{\phi}=(1/r)_{0}{\rm sin}\alpha$ and 
$K^{\theta}=0$, the initial conditions are complete.
A similar approach to the construction of null geodesics on 
spherical structures was used firstly by Panek in 1992.

Finally, we know (see e. g. Anile \& Motta 1976) that the 
CMB temperature measured by the 
observer is related to its value seen at any point $x$ 
along the geodesic specified by $\alpha$ through the relation
\begin{equation}
{T_{x}\over T_{0}}={(K_{\mu}u^{\mu})_{0}
\over (K_{\mu}u^{\mu})_{x}}\ \ ,
\label{Tx/TobsI}
\end{equation}
where $u^{\mu}$ is the four-velocity of detection instrument. 
In the most interesting situation of a void intersecting the LSS, 
a peculiar velocity field causing a Doppler contribution 
exists at emission due to acoustic oscillations of the 
photon-baryon plasma: this contribution, evaluated 
in Sec.(IV) 
together with the dominant or comparable adiabatic effect,
is small with respect to SW one because of Silk damping.

\section{Voids in front of the LSS}

As we already observed, a void,
whether compensated or not, shows an over-comoving 
growth in an ordinary flat Friedmannian background. 
Then, in a first approximation, 
a void can be schematically represented as a region of space 
in which the scale factor $a(t)$ grows faster than $t^{2/3}$; 
more precisely, Bertschinger 
(1983,1985) found $t^{4/5}$ for compensated voids in the
highly non--linear, asymptotic  regime;  our structures, being
mildly non-linear, approximatively obey this law.
Such a dynamics generates a cooling (redshift) 
in the CMB temperature (in addition, of course, to the ordinary one due 
to the unperturbed cosmic expansion). 
\onecolumn
\begin{figure}
\epsfysize=8.0in
\epsfxsize=6.0in 
\epsfbox{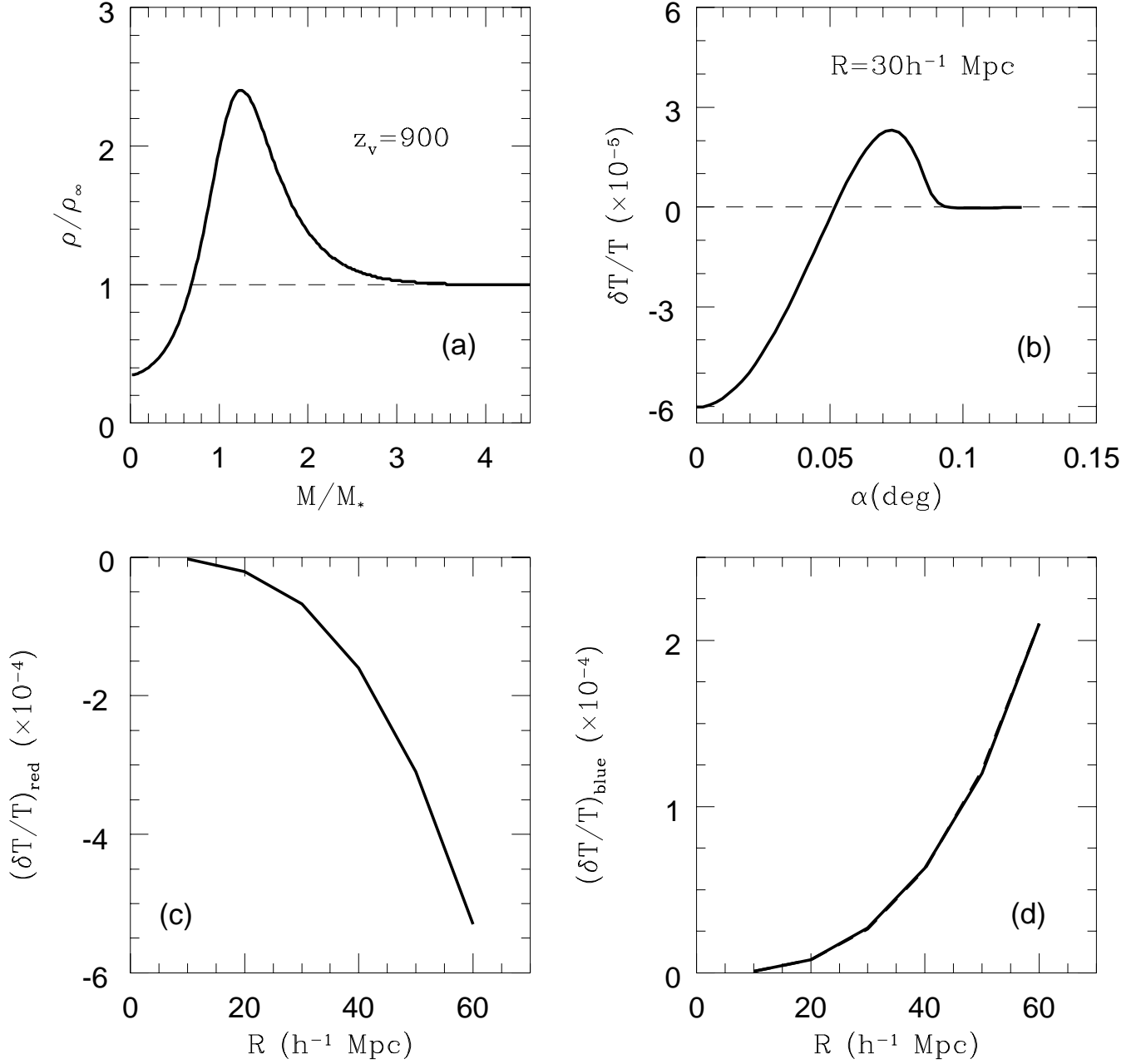}
\caption{RS effect: void between us and the LSS;
complete crossing of the  void by the  CMB photons. The non-linear 
density profile of panel $a$
induces a very characteristic angular dependence on CMB temperature 
(panel $b$). 
In panels $c$ and $d$  dependence on $R$  ($\sim R^{3}$)
of the central redshift  and of the peripheral blueshift.}
\label{fig1}
\end{figure}
\begin{figure}
\epsfysize=8.0in
\epsfxsize=6.0in  
\epsfbox{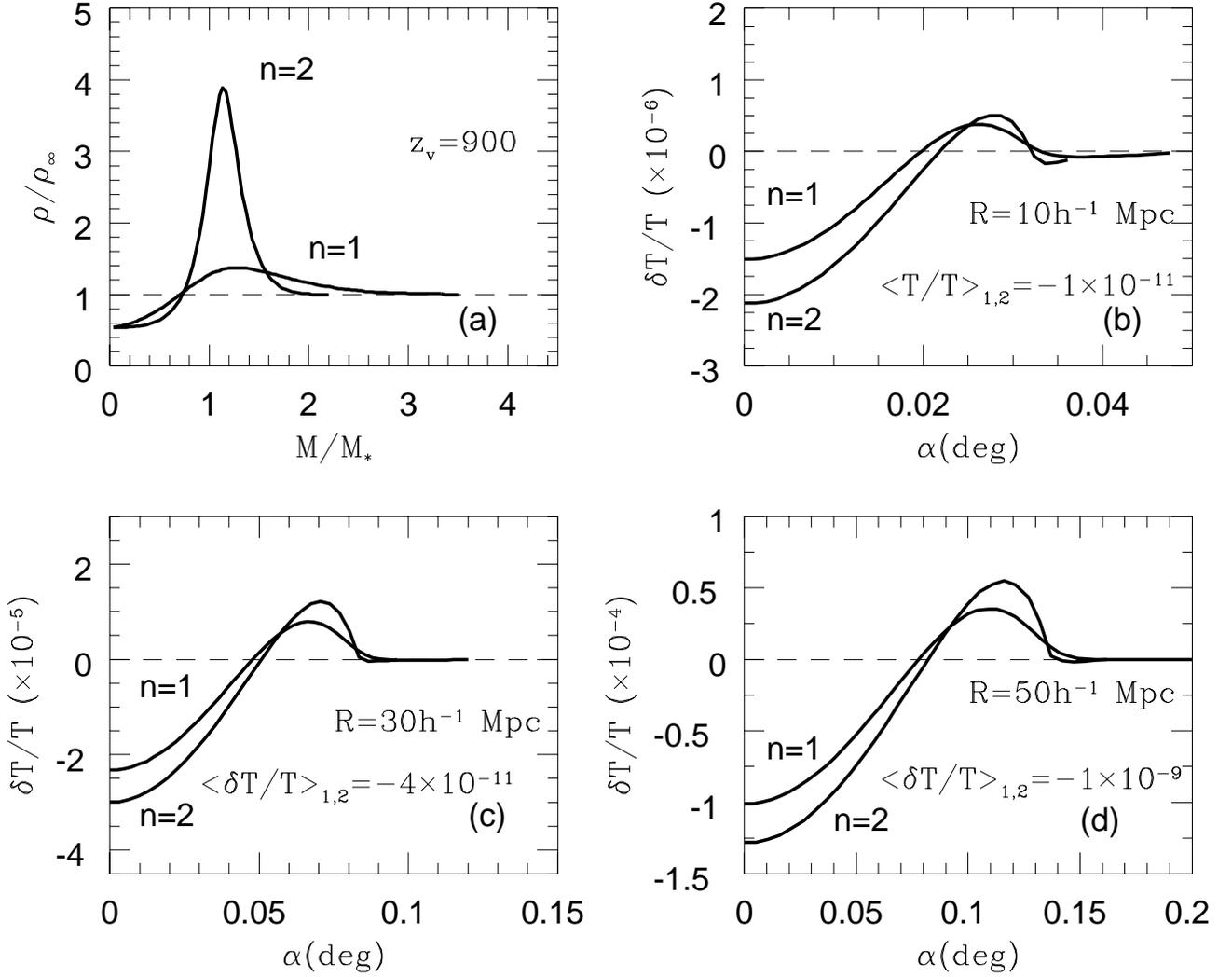}
\caption{RS effect for different values of $n$ (different sharpness of the
density profiles) and for interesting values of the present radii.
The values reported for $<\delta T/T>$, practically indistinguishable for 
$n$=1 and 2, are obtained by simulating 
COBE's angular resolution and appear much below the 
observed anisotropies.}
\label{fig2}
\end{figure}
\twocolumn
On the other hand,
the radiation is heated when crossing the void walls, 
as it is discussed below.

The case to be discussed here concerns a void at $z_{v}= 900$,
 completely crossed by a CMB photon. 
This case was analyzed in TV (firstly) and in recent works (Panek 1992, 
Arnau et al. 1993, Martinez-Gonzales et al. 1990).
We leave to the next section the more important case of 
voids intersecting the LSS. TV found an analytical form
for the angular dependence of the CMB temperature by adopting a 
model for the metric inside a compensated void which, albeit very schematic, 
contains all the essential aspects of the problem.
As it is shown below, our result remarkably agrees with theirs.
Their void is simply a spherical Friedmannian region of 
space in which the scale factor grows in time faster than
in a $\Omega=1$ universe, precisely as $t^{4/5}$.
With an ingenious double Lorentz 
transformation, they find that photons entering and leaving 
the void are boosted by Doppler effect, simply because 
they are seen by frames 
moving towards them; on the contrary, 
during the void's crossing, they cool with respect to the flat 
Friedmannian case, because the space growth is faster.
As the second effect is strongly dependent on the time spent 
by photons inside the void, while the first is
almost fixed, the blueshift becomes dominant
 for strongly off axis trajectories. 
 
Thompson \& Vishniac's analytical result is:
\begin{equation}
{\delta T\over T}=\left({R/(1+z)^{.2}\over H^{-1}}\right)^{3}
\cos\psi_{\alpha}\left({2\over 5}-{2\over 3}\cos^{2}\psi_{\alpha}\right)\ ,
\label{TV}
\end{equation}
where $\psi_{\alpha}$ is the angle between the perpendicular to the shell 
and photon's direction at the leaving point; its expression as a function
of $\alpha$ is 
$$
\psi_{\alpha}=\alpha+
$$
\begin{equation}
+\cos^{-1}\left({d_{v}\sin^{2}\alpha\over R/(1+z)^{.2}}
+\cos\alpha\sqrt{1-{d_{v}^{2}\sin^{2}\alpha\over R^{2}/(1+z)^{.4}}}\right)
\label{psialpha}
\end{equation} 
where $d_{v}$ is the comoving distance of the void's center.

In Fig.(\ref{fig1}) we report our numerical analysis
of a typical void along the line-of-sight ($R=30\hm$   , $z_{v}=900$). 
Panel (a) shows the density profile as a function of the radial 
coordinate $M$ normalized to $M_{*}$, the latter being 
calculated through the value ($\simeq R/4$) of the comoving 
radius at the time of interaction with the CMB.
The central underdensity, obtained for $B=2200$ in
Eq.(\ref{modello}), and the large compensating shell are
well visible.
Panel (b) reports the CMB angular temperature change for complete
crossing; the curve presents all the discussed features and
agrees with the result of Eq. (\ref{TV}).
A completely crossed void appears in the CMB sky as a cold circular spot 
sorrounded by a hot ring, because in off axis  
trajectories the hole contribution is suppressed.
The value predicted by Eq. (\ref{TV}) for central 
trajectories is $-10^{-4}$;  our result is slightly less
because our voids are not completely empty.
The maximal redshift to maximal blueshift
ratio is $-$2.4 in Eq. (\ref{TV}) as in our case. Finally, 
note the smallness of angular sizes, due to the fact that 
this void is at the edge of the observed universe; 
the temperature distortion induced by such a structure is much below the 
experimental limit ($\simeq 10^{-5}$) because the present
instrumentation has a finite angular width 
($3^{o}$ Gaussian beam for COBE).
Panel (c) and (d) show the dependence of maximal redshift 
and blueshift on the void's radius. The solid line is 
numerical, while the dashed one, pratically indistinguishable 
from the first, is $(R/(1+z)^{.2}/H^{-1})^{3}$. 

Now, let us come to the evaluation of the CMB dependence 
on the shell sharpness.
As we stated above, the latter  is regulated by the
parameter $n$. 
We have confronted the CMB temperature changes induced by 
different $n$'s, and  for voids of different sizes in the 
interesting range. The results are shown in Fig.(\ref{fig2}) 
and reveal a very weak dependence on $n$.
The same happens during void crossing. 
In this case (the most interesting),
for any value of the parameter $BL_{s}/L$ in 
Eq. (\ref{modello}), large $n$ makes $\gamma_{n}$ in Eq.(\ref{modellos})
approach the step function; in turn that yields the limiting 
sharpness for the shell's density profile. 
As we found that the variation of $n$ does not change significatively 
the CMB perturbation (Eq. (\ref{calcolo}) below),
we will neglect in the following this effect,
and work with the simplest case $n=1$. 
Also, Fig.(\ref{fig2}) shows $<\delta T/T>$,
defined as a COBE-like Gaussian-averaged signal:
\begin{equation}
<{\delta T\over T}>=N \int {\delta T\over T}(\alpha)
\exp[-\alpha^{2}/2\gamma^{2}] 2\pi\sin\alpha d\alpha\ \ ,
\label{cobe}
\end{equation}
where $\gamma=3^{o}$, COBE's beam size, 
and $N$ is the normalization constant; in the following,
we will refer to the COBE Gaussian as $W(\alpha)$.
It is evident that, as expected,
the values reported
in Fig.(\ref{fig2}), of the order
of $10^{-9}$ for the largest $R$, are extremely small 
compared with the observed anisotropies. 
\onecolumn
\begin{figure}
\epsfysize=8.0in
\epsfxsize=5.5in  
\epsfbox{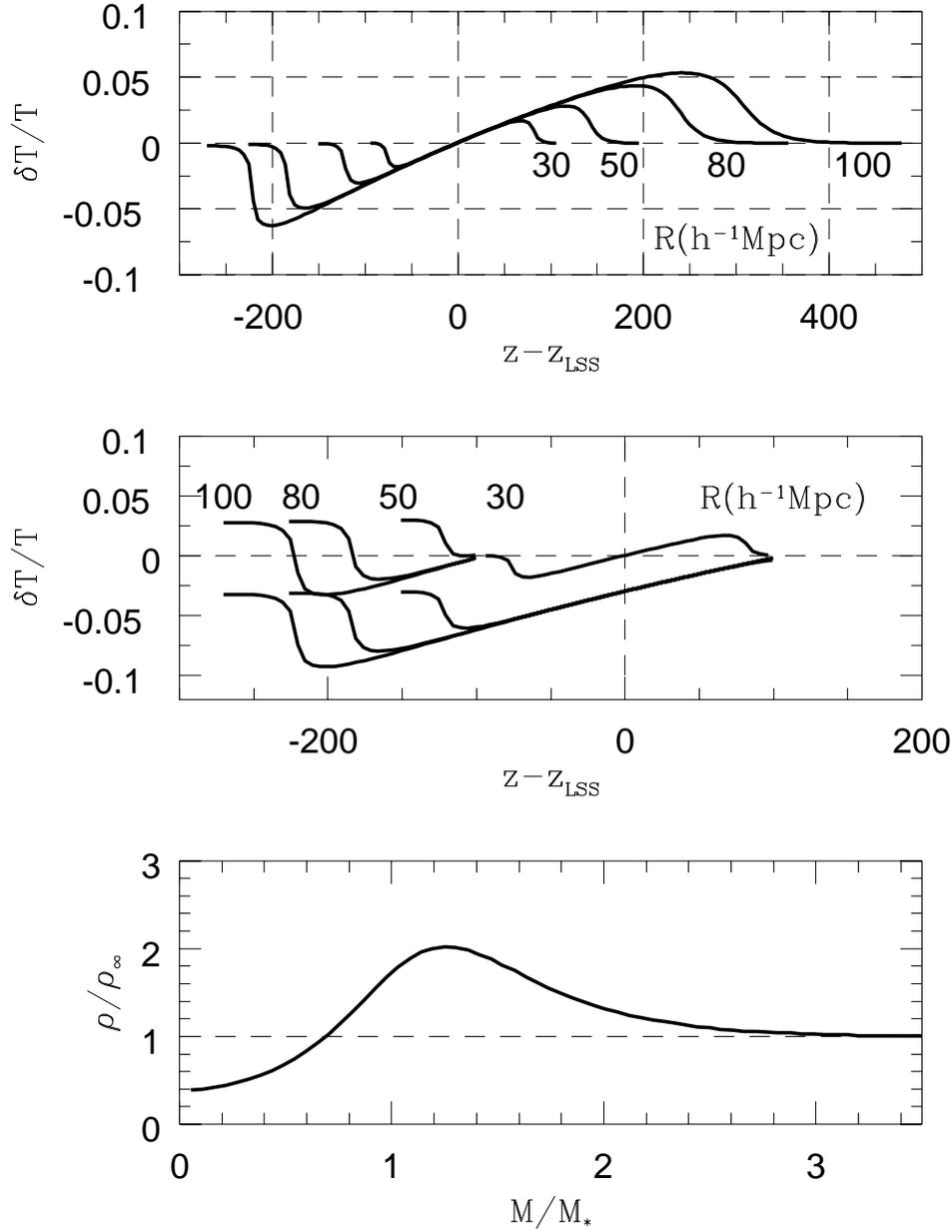}
\caption{Temperature changes followed  as  functions of $z$ along 
 a central trajectory in voids with present physical radii 
$R=$30, 50, 80, $100\hm$  sitting exactly on the LSS.   In panel $a$ the
photons come from infinity with vanishing $\delta T / T$;  realistically,
in panel $b$ they leave either at $z=1100$ or at  $z=900$;
it is then seen that their frequency shift  is either toward the red
or toward the blue depending on the departing position. 
It must be remarked that the order of magnitude of this metric 
induced change is nearly 
10$^{3}$ times its corrispondent value for complete crossing. 
Finally, panel $c$ reports the underlying density profile.}
\label{fig3}
\end{figure}
\twocolumn
We underline that this extreme
smallness derives from the  complete crossing of the void: in fact,
the only observationally interesting 
voids are those intersecting  the LSS, to be analyzed below.
Furthermore, it must also be underlined that voids with $z\ll 1000$ 
will produce an even smaller temperature perturbation, because cosmic 
expansion, not compensated by the over-comoving growth, dilutes them. 

\section{Voids on the LSS}

Fig.(\ref{fig3}) shows the main new result of our 
work. In the upper and in the central panels
 we report the temporal evolution of the 
photon temperature during its crossing of the void: the latter
is shown in the lower panel; 
 different void radii of interest for the observations are shown.
The abscissa is $z-z_{LSS}$, the latter being the redshift of LSS 
($\simeq 1067$); in the four cases shown the void's center lies
exactly on the LSS. In the upper panel 
the photons come from infinity with $\delta T/T=0$
and experience firstly the blueshift from crossing the first shell, 
secondly the strong central 
redshift generated by the over comoving expansion, and finally
the blueshift from crossing the second shell: globally  
this yields the small resulting $\delta T/T$ 
(RS effect) hidden on the left hand border of curves in Fig.(\ref{fig3}). 
This is a very characteristic image of a void crossing, because 
we have found it being nearly the same in form 
for all the trajectories and astrophysically interesting void sizes.
The central panel applies to realistic paths starting from either
edge of the LSS ($z\approx\,900$ and 1100)  and displays the total
blueshifts and redshifts  that can be read on the left hand side.
The order of magnitude of the temperature change during crossing is 
predicted by 
considering the propagation of a photon in a region of space 
expanding as $t^{4/5}$ during a time $\delta t\simeq R$; the result is 
\begin{equation}
\left|{\delta T\over T}\right|\simeq {1\over 5(1+z)^{.2}}{R\over H^{-1}}
\simeq {1\over 20}{R\over H^{-1}}\ \ ,
\label{silk}
\end{equation}
where overcomoving growth is taken into account and $H^{-1}$ is calculated
at interaction epoch $z\simeq 1000$. As may be easily seen from the figure, 
the predicted value is strictly respected;
this argument was inferred firstly by Vadas (1995).
For the reason that $\delta T / T$ is proportional to the time spent
in the void, for non-central trajectories $R$ must be replaced
with $R\sqrt{1-(\alpha /\alpha_{R})^{2}}$, just the void's length 
on a direction characterized by the angle $\alpha$
defined in Sec.(II), the maximum value of which is $\alpha_{R}$.
An accurate analytic approximation of the results for all 
trajectories is 
\begin{equation}
{\delta T\over T} = {R\, \sqrt{1-(\alpha /\alpha_{R})^{2}}\over 20
\, H^{-1}}
\, 1.25\,x\exp\left[-\left({x\over 1.19}\right)^{10}\right] 
\ \ ,
\label{bacci}
\end{equation}
where 
\begin{equation}
x={r\over \sqrt{1-(\alpha /\alpha_{R})^{2}}R/4}
\label{baccix}
\end{equation}
and $r$ is the photon's coordinate.
We see that the sharpness is already very strong in the present case $n=1$; 
increasing $n$ up to infinity simply brings the curves in Fig.(\ref{fig3})
to their limiting case
\begin{equation}
{\delta T\over T} = \, {R\, \sqrt{1-(\alpha /\alpha_{R})^{2}}\over 20
\, H^{-1}}\, 1.25\, x\, 
\theta(|x|-1)\ \ , 
\label{baccilimit}
\end{equation}
where $\theta$ is the Heaviside function.
We also see that an axial
trajectory presents a typical  $\delta T / T$
{\it three orders of magnitude above} the central
one of Fig.(\ref{fig1}) (panel (b)), obtained 
for complete crossing and hidden on the left hand border 
of curves of the RS effect, displayed in the upper panel of Fig.(\ref{fig3}).
The reason why this strong contribution does not appear in Eq. (\ref{TV})
is that for complete crossing the effects of the shells and the central
hole balance out, as it is evident from Fig.(\ref{fig3}), leaving
place to terms with higher power of $R/H^{-1}$.
Therefore, if a void intersects the LSS,
the ensuing $\delta T / T$ may be very different depending on the particular 
LSS position with respect to the center of the void. In any case,
the amplitude will be much larger than in the case of complete crossing.
Let us make a simple calculation: an useful approximation
to the total temperature distortion induced by a void
of angular size $\alpha_{R}$ averaged over a beam of size $\gamma$ is
\begin{equation}
<{\delta T\over T}>\simeq \left({\delta T\over T}\right)_{\alpha=0}
{\alpha_{R}^{2}\over \gamma^{2}}\ \ ,
\label{cobeapp}
\end{equation}
and taking $\delta T/T |_{\alpha=0}\simeq
10^{-2}$ as a typical value from Fig.(\ref{fig3}) we obtain for 
$R=100\hm$  ($R=25\hm$ at decoupling) and $\gamma =3^{o}$
\begin{equation}
<{\delta T\over T}>\simeq 6\cdot 10^{-5}\ \ .
\label{cobelss}
\end{equation}
This value, barely above the present observations, indicates that 
such scales are very close to the maximum permitted 
to primordial voids from CMB isotropy. This motivates  
the accurate investigation below. 
\onecolumn  
\begin{figure}
\epsfysize=8.0in 
\epsfxsize=6.0in
\epsfbox{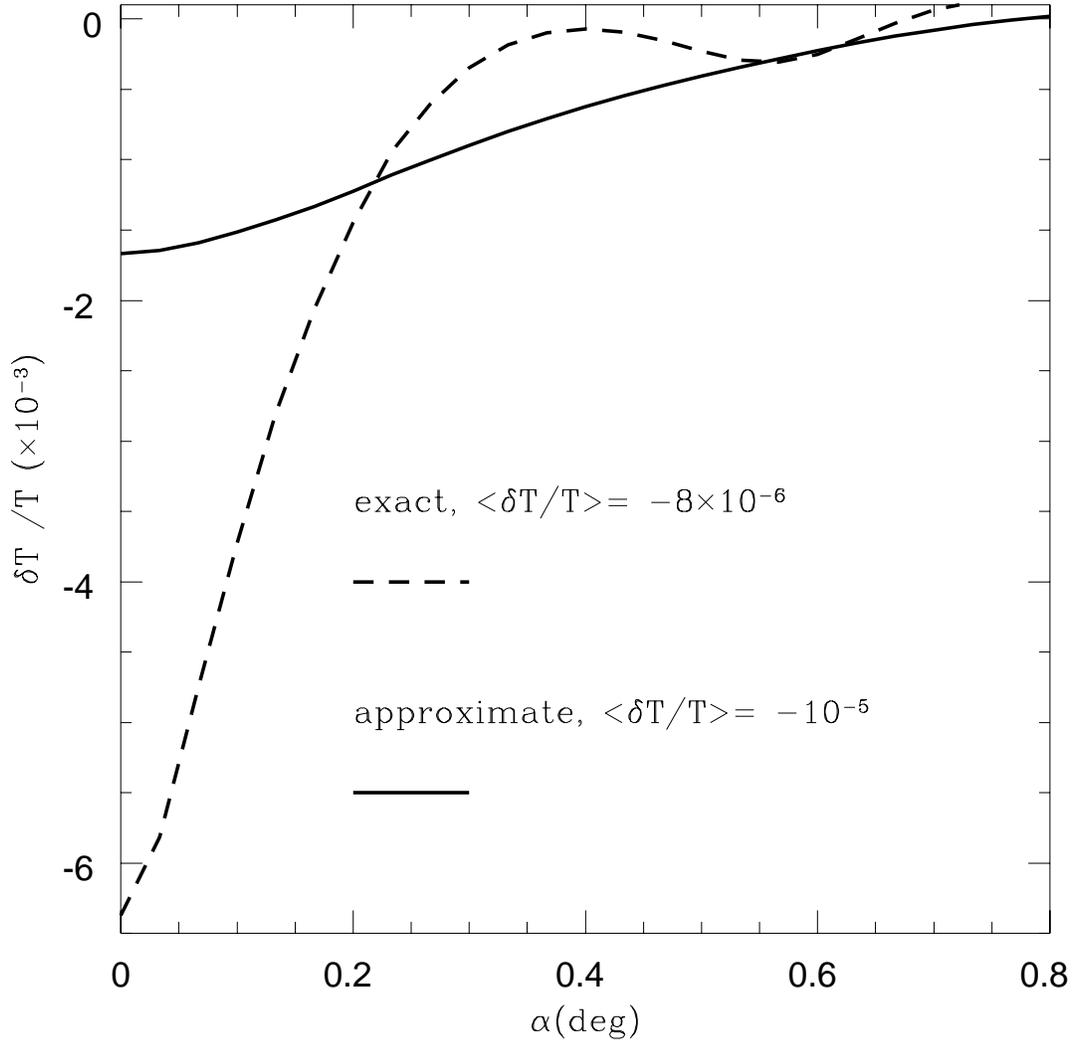}
\caption{Purely adiabatic and Doppler $\delta T / T$
due to a compensated void in the cosmic fluid, both in the 
approximate (solid curve) and exact (dashed curve) case. 
On the relevant scales, this perturbation is well below
the SW effect.}
\label{fig4}
\end{figure}
\twocolumn

To treat accurately the case of voids intersecting the LSS, we
must take care of all the physical processes at work.
In fact the SW effect on the photons decoupled inside a void is
not the only perturbation to the CMB; the corresponding inhomogeneity
of the photon-baryon plasma naturally
induces adiabatic and Doppler temperature fluctuations before decoupling.
We anticipate that such contributions are reduced
by Silk damping on scales $\le 25\hm$ in which
the SW effect is already comparable with the observed CMB anisotropy.
To see this, we have used the formalism developed by HS; 
it allows to obtain the CMB
fluctuations at decoupling caused by a perturbation characterized 
by a Newtonian potential $\phi$ (our perturbations are not properly
in the linear regime, but this approximation allow us to estimate
correctly the effect). 
In the HS treatment, the Boltzmann, Euler and continuity 
equations for the photon-baryon plasma,
accounting for Thomson diffusion together with gravitational and pressure
terms, are analytically solved
under the tight-coupling approximation. In our treatment, in order 
to estimate only the adiabatic and Doppler effects, we have excluded
the SW terms from HS formulas; we performed the calculus exactly and
with the following simplifications:
$a)$ constant gravitational potentials, which is nearly true in the 
linear regime 
and approximatively in our mildly non-linear case;
$b)$ the quantity $3\rho_{b}/4\rho_{\gamma}$, that goes from 0
to .1 from the remote past to decoupling in models with 
$\Omega_{b}\simeq .01$, is neglected (equivalent to have a 
constant sound speed); $c)$ istantaneous decoupling, and
the Silk Damping scale taken  at this time.
With these hypotheses in the HS treatment,
the amplitude of Fourier {\bf k}-mode of temperature  
fluctuation at decoupling $(_{D})$ (we report only the monopole term
for simplicity) is related to the potential 
by the simple and intuitive relation
\begin{equation}
\left({\delta T\over T}\right)_{k,D}=\phi_{k,D}\left(1-{1\over 2}\cos
kc_{s}\eta_{D}\right)e^{-(k/k_{D})^2}\ \ ,
\label{hs}
\end{equation}
where $\lambda_{D}=2\pi /k_{D}=
3(\Omega_{b}h^{2})^{-1/2})(\Omega_{o}h^{2})^{-1/4}$ Mpc is the 
Silk damping scale at decoupling
(again from HS formulas),
and $\eta$ is the conformal time. 
On scales larger than
the sound horizon $c_{s}\eta_{D}$, Eq.(\ref{hs}) reproduces the
known behaviour 
$\delta T/T\propto\phi$ up to a factor of the order of unity;
on scales smaller than 1/$k_{D}$
Silk damping forces to zero the inhomogeneities of the photon-baryon
plasma. At intermediate scales, corresponding to $.2^{o}\div 1^{o}$ in the
sky, an oscillatory behaviour occurs due to the acoustic oscillations;
it will be retained in the form of Doppler peaks in the correlation
function of $\delta T/T$.
As in HS, one can antitransform $(\delta T /T)_k$     
to obtain the contribution to
$\delta T/T$ from the adiabatic effect as a function of the angle 
$\alpha$. For the
potential source we have used the simple model
\begin{eqnarray}
{\delta\rho\over\rho}(r)&=&-\delta\,, \quad 
{r\over R}\le 1\ \ ,\nonumber \\
{\delta\rho\over\rho}(r)&=&{\delta\over 
(1+a)^{3}-1}\,, \quad
	1\le {r\over R}\le 1+a\,,\nonumber \\
{\delta\rho\over\rho}(r)&=&0\,,\quad {r\over R}\ge 1+a\,,
\label{DRR} 
\end{eqnarray}
i.e. a central cavity of depth $1-\delta$ surrounded by a compensating 
thin shell $\Delta R=a\cdot R$.
The result, together with the exact calculus,
is shown in Fig.(\ref{fig4}) for the interesting values
$(1+z_{D})^{.2}R=25\hm$, a=.5, $\delta=.7$ and the typical set
$\Omega_{b}=.06,\Omega_{0}=1,h=.5$ for a CDM universe; 
the signal that COBE would detect for this 
perturbation is also reported and proves to be below 
the SW contribution, as we have anticipated.

Therefore, the picture of a primordial void sitting on the LSS which
emerges very clearly from this analysis is that of a strong spot of metric 
origin and of either color on very small angular scales.

For the above arguments, we are also allowed to
use the standard homogeneous model for the width of
LSS. By taking into account 
the dependence of the fractional ionization on redshift $z$, and its 
consequences upon optical depth for photons,
we are led to the probability function for a photon to be last scattered
at redshift $z$ (Jones and Wyse, 1985)
\begin{equation}
P(z)={5.26\over 1000}\left({z\over 1000}\right)^{13.25}
\exp\left[-0.37\left({z\over 1000}\right)^{14.25}\right]
\label{P(z)}
\end{equation}
that is well approximated by a Gaussian centered in $z_{max}\simeq 1067$
and with $\sigma\simeq 90$.

Now, let's use current observational data on the CMB isotropy to 
estabilish constraints on voids at decoupling.
As we stated above, the qualitative form of the curves in 
Fig.(\ref{fig3}) is the same 
regardless of the voids' radii. The COBE observed anisotropies distribution 
have an amplitude $\sigma\simeq 3\cdot 10^{-5}$ 
at small angular scales (Bennett $et\ al.$ 1994): therefore we will assume
that a void is excluded by current observational data if its $<\delta T / T>$
is $3\sigma\simeq 10^{-4}$.
\onecolumn
\begin{figure}
\epsfysize=8.0in 
\epsfxsize=6.0in
\epsfbox{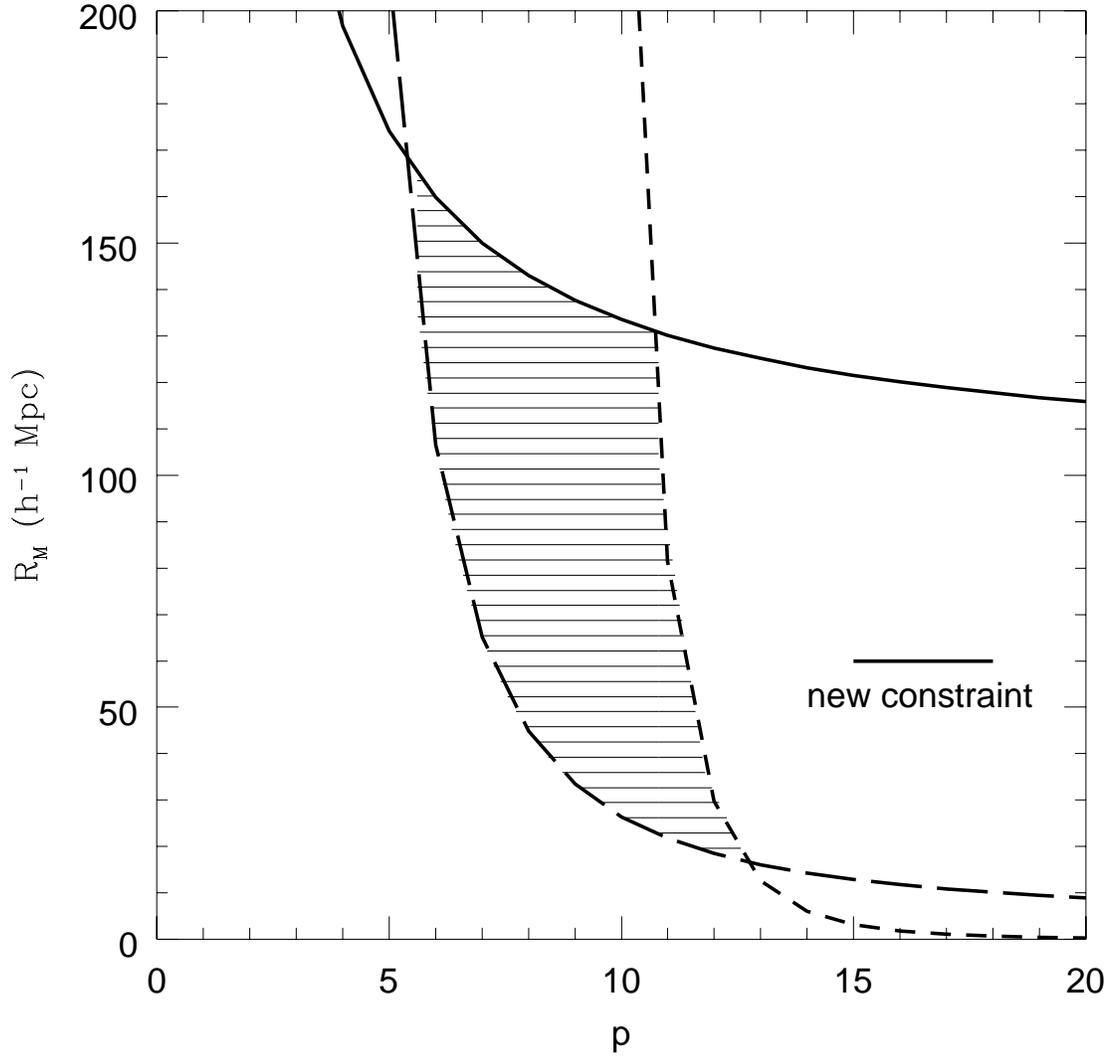}
\caption{Observational constraints on the parameters $p,R_{M}$:
the long-dashed line refers to the request that a reasonable fraction
of space (50\% )
is filled by primordial voids; the short-dashed line is based on small 
hot spots statistics; the solid line is the present result: 
the number of voids inducing a visible CMB perturbation must 
be smaller than one. The shaded region
gives the acceptable parameter space for primordial void 
distribution in the whole observed universe.}
\label{fig5}
\end{figure}
\twocolumn 
We have integrated the CMB fluctuation for many void radii in the
relevant range, to get $<\delta T/T>$  as a function of $R$:
\begin{equation}
<{\delta T\over T}(R)>=\int 2\pi \sin\alpha d\alpha dz 
{\delta T\over T}(R,\alpha,z)\, W(\alpha)P(z)\ .
\label{calcolo}
\end{equation}
We can immediately obtain an upper limit for $R$ as follows. 
At radial distance $(1+z)^{.2}r_{peak}=R$
inside the void, the CMB distortion, whether red
or blue, reaches its maximal value before going to zero
(see Fig.(\ref{fig3})).
If $r_{peak}$ coincides with
the peak in $P(z)$
(at $z_{LSS}=1067$), the CMB distortion
detected by an
instrument with Gaussian window function with width $\gamma$ 
(expressed in degrees) is found to be
\begin{equation}
<\left|{\delta T\over T}\right|>\simeq {10^{-3}\over\gamma^{2}}
\left({R\over H^{-1}}\right)^{3}\ \ ,
\label{fit}
\end{equation}
where the numerical factor contains a correction 1/3 due to the LSS 
width for $R$ near to $H^{-1}$.
An $R$ comes from the SW effect, while the remaining $R^{2}$ is
due to the beam's angular width. Then,
COBE-like experiments ($\gamma=3^{o}$) say that
no voids with $R = .97\, H^{-1}\simeq 100\hm$ (25 at decoupling)
or more exist on the LSS.
 We point out that
the same result without considering the LSS width 
would be $R=\, .63\, H^{-1}$.

Now that we have the largest void passing the CMB bounds,
we pass to evaluate the statistical consequences on
distribution (\ref{distribution}). 
The position of the LSS 
peak with respect to the void center corresponds to a value of 
$x$ defined in Eq.(\ref{baccix}); for any $x$, there will
be a radius $R_{max}(x)$ for which the CMB distortion detected 
by COBE would be nearly $3\sigma$; this is obtained numerically
from Eq.(\ref{calcolo}), putting the LSS peak at $r=x\, R$
from the void's center. 
The number of voids with comoving radius $R$ is 
$|dN_{B}/dR|dR$, and the fraction of them having the LSS peak 
at distance between $r$ and $r+dr$ is
\begin{equation}
dN(R,x)=\left|{dN_{B}\over dR}\right|dR{3\, dr\over 2H_{o}^{-1}}=
\left|dN_{B}\over dR\right|\, R\, dR {3\, dx\over 2H_{o}^{-1}}\ \ .
\label{dnRx}
\end{equation}
The total number of such voids with $R\ge R_{max}$ must be 
less than 1; this is equivalent to the condition
\begin{equation}
\int_{-\infty}^{+\infty}dx\int_{R_{max}(x)}^{+\infty}dR
\left|{dN_{B}\over dR}\right|{3\, R\over 2H_{o}^{-1}}<1\ \ .
\label{intdnRx}
\end{equation}
By using Eq.(\ref{distribution}) we
obtain the following upper limit relation between $R_{M}$ and $p$:
\begin{equation}
R_{M}<\left({p\over p-1}\int_{-\infty}^{+\infty}R_{max}(x)^{1-p}
{3\, dx\over 2H_{o}^{-1}}\right)^{-1/p}\ \ .
\label{constraint}
\end{equation}
The result is shown in Fig.(\ref{fig5}) as a solid line; 
values above 
the long-dashed 
line ensures that at least
half of the space is presently filled by large scale 
voids ($R\ge 10\hm$   ); values below
the short-dashed line 
ensure that the random noise produced by an overposition of the
many small voids on the CMB does not break the CMB isotropy
(Amendola \& Occhionero 1993). 
The dashed area contains roughly the set of $(p,R_{M})$ values 
\begin{equation}
6\le p\le 13\ \ ,\ \ 30\hm\   \le R_{M}\le 130\hm\   \ ,
\label{result}
\end{equation} 
that is the parameter space for which the voids
respect all the CMB constraints worked out so far,
and provide large scale power matching current observations
of galaxy clustering.

\section{Conclusions}

We have built a formalism that allows us to check the plausibility of the 
cosmogonic role of primordial voids; here, strictly speaking, by 
``primordial" we mean ``present already at decoupling": however
the obvious suggestion is that the voids originate much earlier,
precisely in an inflationary first order phase transition. The
constraints from the observations lie entirely in the compatibility
of the voids' theoretical  imprints on the CMB with the observed
anisotropies at the relevant scales and translate into defining
an allowed region of the parameter space for the void distribution.

In order to compute photon trajectories and temperature perturbations,
we describe a void  by an appropriate analytic general relativistic
model of a spherical perturbation embedded in a flat Friedmann background.

Firstly we remind from the literature the case of a void lying between us and 
the last scattering surface (LSS): this implies the  complete photon
crossing known as Rees-Sciama effect. It is known that this induces
a $\delta T / T \simeq -(4/15)(R/(1+z)^{.2}H^{-1})^3 $: we have found
that this results from a partial cancelation of the redshift occurring
during the void crossing, namely the  Sachs-Wolfe effect,
with the blueshifts occurring at either
shell crossing. Each of these effects,
$|\delta T / T |\simeq R/20H^{-1} $,
 is three orders of magnitude larger than the total effect.
 
 In the present paper, we have therefore considered the case of voids
 lying on the LSS and we have evaluated the global effect 
 for photons decoupled inside the void, comprising the Sachs-Wolfe
 contribution mentioned above, the adiabatic and Doppler effects
 due to the acoustical oscillations of the photon-baryon plasma;
 the latter has been found small with respect to the former
 because Silk damping is active on the small scales involved.
 For completeness, we have used an exact model for the LSS profile.
 Therefore, we are led to a very clear picture for the CMB impact of
 primordial voids: they are strong spots of metrical origin on 
 very small angular scale.
 
 A first observational constraint is obtained by requiring that 
 the number of astrophysically interesting
 voids intersecting the LSS and capable of generating
 a distortion above the present observational limits on small angular 
scales, is smaller than unity. In practice this yields the available 
region of the parameter space for the void distribution, which we choose
in the convenient form of a power law.

Where do we go from here? We expect a lot of power on the very small
angular scales, $\ell \simeq 1000$ in the canonical expansion in Legendre
polynomials of the CMB correlation function. Power on larger scales,
$\ell \le 100$, comes from the ordinary zero point fluctuations of the 
inflaton as well as from the variation of the number of voids in different 
regions of the sky,
while in our case an unexpected strong correlation of metric origin 
arises on scales where
Silk damping erases everything else. We plan therefore to build realistic
sky maps by taking into account all  the voids on the LSS in order
to recognize whether 
the imprint of primordial voids will be actually present
in the data of future high resolution experiments 
like $COBRAS-SAMBA$. 
  
We thank David Wands and Michael Sahzin for useful comments.

\end{document}